\documentstyle[emulateapj,psfig]{article}
\onecolumn
\singlespace
\newcommand{\cmm}{\,{\rm cm}^{-2}}

\newcommand{\Lya}{Ly$\alpha\ $}

\newcommand{\etal}{et~al.\ }

\def\kms{\,{\rm km\,s^{-1}}}
\def\kmsmpc{\,{\rm km\,s^{-1}\,Mpc^{-1}}}
\def\msun{\,{\rm M_\odot}}

\def\spose#1{\hbox to 0pt{#1\hss}}
\def\lta{\mathrel{\spose{\lower 3pt\hbox{$\mathchar"218$}} \raise 2.0pt\hbox{$\mathchar"13C$}}}
\def\gta{\mathrel{\spose{\lower 3pt\hbox{$\mathchar"218$}} \raise 2.0pt\hbox{$\mathchar"13E$}}}
\def\cmm{\,{\rm cm}^{-2}}

\def\lya{Ly$\alpha\ $}
\def\uvunits{{\rm\,erg\,cm^{-2}\,s^{-1}\,Hz^{-1}\,sr^{-1}}}
\def\units{{\rm\,erg\,cm^{-2}\,s^{-1}\,Hz^{-1}}}

\def\ni{\noindent}
\def\HI{\hbox{H~$\scriptstyle\rm I\ $}}
\def\HII{\hbox{H~$\scriptstyle\rm II\ $}}

\def\HeII{\hbox{He~$\scriptstyle\rm II\ $}}

\def\CIV{\hbox{C~$\scriptstyle\rm IV\ $}}
\def\SIV{\hbox{Si~$\scriptstyle\rm IV\ $}}
\def\OVI{\hbox{O~$\scriptstyle\rm VI\ $}}
\def\nHI{{\rm HI}}
\def\nH{{\rm H}}

\def\vir{{\rm vir}}
\def\igm{{\rm IGM}}
\def\gtsima{$\; \buildrel > \over \sim \;$}
\def\ltsima{$\; \buildrel < \over \sim \;$}
\def\prosima{$\; \buildrel \propto \over \sim \;$}
\def\gsim{\lower.5ex\hbox{\gtsima}}
\def\lsim{\lower.5ex\hbox{\ltsima}}
\def\simgt{\lower.5ex\hbox{\gtsima}}
\def\simlt{\lower.5ex\hbox{\ltsima}}
\def\simpr{\lower.5ex\hbox{\prosima}}

\def\ie{{\frenchspacing i.e. }}

\newcommand\beq{\begin{equation}}
\newcommand\eeq{\end{equation}}

\lefthead{Madau, Ferrara \& Rees}
\righthead{IGM Metal Enrichment}
\submitted{ApJ, in press}
\begin{document}

\title{Early Metal Enrichment of the Intergalactic Medium by Pregalactic 
Outflows}
\author{Piero Madau\altaffilmark{1,2,3}, Andrea Ferrara\altaffilmark{2,4}, and 
Martin J. Rees\altaffilmark{1}}

\altaffiltext{1}{Institute of Astronomy, Madingley Road, Cambridge CB3 0HA, UK.}
\altaffiltext{2}{Osservatorio Astrofisico di Arcetri, Largo E. Fermi 5, 50125 Firenze, Italy.}
\altaffiltext{3}{Department of Astronomy and Astrophysics, University of
California, Santa Cruz, CA 95064.}
\altaffiltext{4}{Center for Computational Physics, University of Tsukuba, 
Tsukuba--shi, Ibaraki--ken, 305--8577, Japan.}

\begin{abstract}
We assess supernova (SN)--driven pregalactic outflows 
as a mechanism for distributing the product of stellar nucleosynthesis over 
large cosmological volumes prior to the reionization epoch. 
SN ejecta will escape the grasp of 
halos with virial temperatures $T_\vir\gta 10^{4.3}\,$K (corresponding to masses
$M\gta 10^8\,h^{-1}\,\msun$ at redshift $z=9$ when they collapse from 2--$\sigma$ 
fluctuations) if rapid cooling can take place, and a significant fraction of 
their baryonic mass is converted into stars over a dynamical timescale. 
We study the evolution of SN--driven bubbles as they blow out from subgalactic
halos and propagate into the intergalactic medium (IGM), and show that to lift the
halo gas out of the potential well the energy injection must continue 
at least until blow--away occurs.
If the fraction of ionizing photons that escape the dense sites of star formation
into the intergalactic space is greater than a few percent, pregalactic outflows
will propagate into an IGM which has been pre--photoionized by the same massive 
stars which later explode as SNe, and the expansion of the metal--enriched 
bubbles will be halted by the combined action of external pressure, gravity, and 
radiative losses. The collective explosive output 
of about ten thousands SNe per $M\gta 10^8\,h^{-1}\,\msun$ halo at these early 
epochs could pollute vast regions of intergalactic space to a mean metallicity 
$\langle Z\rangle=\Omega_Z/\Omega_b\gta 0.003$ (comparable to the levels 
observed in the \Lya forest at $z\approx 3$) without much perturbing the IGM 
hydrodynamically, i.e. producing large variations of the baryons relative 
to the dark matter. Rayleigh-Taylor instabilities between the 
dense shell which contains pristine swept--up material and the hot, 
metal--enriched
low--density bubble may contribute to the mixing and diffusion of heavy 
elements. The volume filling factor of the ejecta 
is higher than 20\% if the star formation efficiency is of order 10\%. 
Larger filling factors (not required by current observations) may be obtained 
for larger efficiencies, moderately
top--heavy IMFs, halos where a significant fraction of the gas is 
in a galactic disk and does not couple to the outflow (as matter is ejected 
perpendicularly to the disk), or from a population of more numerous sources -- 
which would therefore have to originate from lower amplitude peaks.  When the 
filling factor of the ejecta becomes significant, enriched 
material will typically be at a higher adiabat than expected from photoionization.

\end{abstract}
 
\section{Introduction}

In currently popular cosmological scenarios -- all variants of the cold dark 
matter (CDM) cosmogony with different choices for the parameters $\Omega_M,
\Omega_\Lambda, \Omega_b, h, \sigma_{8}, n$ -- some time beyond a redshift of 
15 the gas 
within halos with virial temperatures $T_\vir\gta 10^4\,$K [or, equivalently, 
$M\gta 10^9\, (1+z)^{-3/2}h^{-1}\,\msun$]\footnote{Throughout this paper we 
will assume, unless stated 
otherwise, an Einstein--de Sitter (EdS) universe with $H_0=100\,h\,\kmsmpc$.}\, 
cooled rapidly due to the excitation of 
hydrogen \Lya by the Maxwellian tail of the electron distribution, and 
fragmented. Massive stars formed with some initial mass function (IMF), 
synthesized heavy elements, and exploded as 
Type II supernovae (SNe) after a few$\times 10^7\,$yr, enriching the 
surrounding medium; these subgalactic stellar systems, aided perhaps by an 
early population of accreting black holes in their nuclei, generated the 
ultraviolet radiation and mechanical 
energy that reheated and reionized the universe. While collisional excitation 
of molecular hydrogen may have allowed the gas in even smaller systems [virial 
temperatures of only a few hundred K, corresponding to masses around 
$10^7\,(1+z)^{-3/2}h^{-1}\,\msun$] to cool and form stars at earlier times 
(Couchman \& Rees 1986; Haiman, Rees, \& Loeb 1996; Tegmark \etal 1997), H$_2$ 
molecules are 
efficiently photo--dissociated by stellar UV radiation, and such `negative 
feedback'  could have suppressed molecular cooling and further star 
formation inside very small halos (e.g. Haiman, Abel, \& Rees 2000; 
Ciardi, Ferrara, \& Abel 2000).  

Throughout these crucial formative stages, the all--pervading intergalactic 
medium (IGM) acted as a source for the gas that is accreted, cools, and 
forms stars within these subgalactic systems, and as a sink for the 
metal--enriched material, energy, and radiation which they eject. The 
well--established 
existence of heavy elements like carbon, nitrogen, and silicon in the \lya 
forest clouds at $z=3-3.5$ may be the best evidence for such an 
early episode of pregalactic star formation. The detection of weak but 
measurable \CIV and \SIV absorption lines in clouds with \HI column densities 
as low as $10^{14.5}\,\cmm$ implies a minimum universal metallicity relative
to solar in the range $[-3.2]$ to $[-2.5]$ at $z=3-3.5$ (Songaila 1997). 
There is no indication in the data of a turnover in the \CIV column density 
distribution down to $N_{\rm CIV}\approx 10^{11.7}\, \cmm$ ($N_\nHI\approx 
10^{14.2}\,\cmm$, Ellison \etal 2000); the analysis of individual pixel 
optical depths may actually imply the presence of weak metal lines below the 
detection threshold (Cowie \& Songaila 1998; Ellison \etal 2000). 
Widespread enrichment is further supported by the recent observations 
of \OVI absorption in low density regions of the IGM (Schaye \etal 2000).

In this paper we argue that the observed paucity of regions in the IGM which 
are of truly primordial composition -- or have abundances as low as those of 
the most metal--poor stars in the Wilky Way halo (Ryan, Norris, \& Beers 
1996) -- may point to an early enrichment epoch by low--mass subgalactic 
systems, rather than being due to late pollution by massive galaxies.
Whilst outflows of metal--rich gas are direcly observed in local 
starbursts (e.g. Heckman 1999) and $z\approx 3$ Lyman--break galaxies 
(Pettini \etal 2000), most of this gas may not leave these massive galaxies 
altogether, but remains trapped in their gravitational 
potential wells  until it cools and rains back onto the galaxies.
By contrast, metal--enriched material from SN ejecta is far more easily 
accelerated to velocities larger than the escape speed -- about $80\,\kms$ 
at the center of a $10^8 h^{-1}\,\msun$  Navarro--Frenk--White (1997, hereafter NFW) 
halo at $z=9$ -- in the shallow potential wells of subgalactic systems (Larson 1974;
Dekel \& Silk 1986).
These early protogalaxies -- with masses comparable to those of 
present--day dwarf ellipticals -- are then expected to release significant
amounts of kinetic energy, heat, and heavy elements into the surrounding 
intergalactic gas. Many authors have addressed the impact of SN--driven winds 
from small starbursting galaxies at high--$z$ (e.g. Barkana \& Loeb 2001;
Ferrara, Pettini, \& Shchekinov 2000; Scannapieco \& Broadhurst 2001; MacLow 
\& Ferrara 1999; 
Murakami \& Babul 1999; Nath \& Trentham 1997; Voit 1996; Tegmark, Silk, \& 
Evrard 1993) or from massive galaxies at more modest redshifts (Aguirre 
\etal 2000; Theuns, Mo, \& Schaye 2001) on the thermal and chemical state of 
the IGM. Gnedin \& Ostriker (1997) and Gnedin (1998) have argued that violent 
merging between protogalaxies 
may be an alternative mechanism to transport metals in the low density IGM,
while Barkana \& Loeb (1999) have discussed the possibility that 
enriched material in halos with virial temperatures $\lta 10^4\,$K 
could be photoevaporated at reionization. Here we give an idealized 
assessement of pregalactic outflows as a mechanism for 
distributing the products of stellar nucleosynthesis over large volumes. 
We will be concerned exclusively with explosive multi--SN events
operating on the characteristic timescale of a few $\times 10^7\,$ yr, 
the lifetime of massive stars. In a 
complementary study, Efstathiou (2000) has recently shown 
that a large fraction
of the baryonic mass of galaxies with virial temperatures $\approx 10^5\,$K 
can also be expelled in `quiescent' mode, i.e. over the relatively long timescale 
of 1 Gyr. To anticipate the conclusions of this work, we find that 
the IGM will be polluted over large scales at the end of the `dark ages' if a 
fraction greater than a few percent 
of the baryonic mass of subgalactic halos can be converted into stars over a 
dynamical timescale. In the case of large star formation efficiencies and/or 
moderately top--heavy IMFs, the temperature of vast regions of the IGM will be 
driven to a higher adiabat
than expected from photoionization, so as to 
inhibit in these regions the formation of further protogalaxies 
by raising the Jeans mass.

\section{An early enrichment epoch?}

Before embarking on a discussion of the consequences of pregalactic outflows
on the thermal and chemical state of the IGM, it is worth summarizing a few
key observational facts and theoretical results:

\ni (1) Numerical N--body/hydrodynamics simulations of structure formation in
the IGM within the framework of CDM dominated cosmologies (e.g. Cen \etal 1994; 
Zhang, Anninos, \& Norman 1995; Hernquist \etal 1996; Theuns \etal 1998) have
recently provided a coherent picture for the origin of the \Lya forest, one 
of an interconnected network of sheets and filaments with virialized systems 
(halos) located at their points of intersection. 
The simulations show good agreement with the observed line
statistics under the assumption that an IGM with baryon density parameter
$\Omega_bh^2=0.019$ (Burles and Tytler 1998) is photoionized and photoheated by a UV
background with a hydrogen ionization rate $\Gamma(z=3.5)\approx 0.6\times 10^{-12}
\,$s$^{-1}$, close to that inferred from quasars (Haardt \& Madau 1996). They also 
show a clear correlation between \HI column, gas temperature, and overdensity, with
\beq
\rho_b/\overline\rho_b\simeq 0.8\,N_{\nHI,13}^{0.7}~~~~~~~{\rm 
and}~~~~~~~
\rho_b/\overline\rho_b\simeq 0.3\,T_4^2  \label{eq:sim}
\eeq
at $z=3.5$ (Ricotti, Gnedin, \& 
Shull 2000; Zhang \etal 1998; Hui \& Gnedin 1997).\footnote{
Throughout this work we adopt the notation $Y_x=Y/10^x$, and cgs units.}\,
In photoionization equilibrium, an optically thin cloud with internal 
density $\rho_b$ will have a neutral hydrogen fraction of 
\beq
{n_\nHI\over n_\nH}\simeq 10^{-5}\,{\rho_b\over {\overline\rho_b}}\,
\left({1+z\over 4.5}\right)^3 T_4^{-0.7}\,\Gamma^{-1}_{-12.2}, \label{eq:nf}
\eeq
where the temperature dependence is from the radiative recombination rate.
Combining equations (\ref{eq:sim}) and (\ref{eq:nf}), and omitting for 
simplicity the redshift and photoionization rate scalings, one obtains 
approximately $n_\nHI/n_\nH\simeq 
10^{-5.3}\,N_{\nHI,13}^{1/2}$. The baryonic mass fraction in the \lya forest per
unit logarithmic \HI column interval can be written as
\beq
f_m={1.3m_p\over \Omega_b\rho_{\rm crit}}\,{N_\nHI^2 f(N_\nHI,z)\over n_\nHI/
n_\nH}\,{dz\over cdt},
\eeq
where $m_p$ is the proton mass, $\rho_{\rm crit}=3H^2/8\pi G$ is the critical 
density of the universe at redshift $z$, $f(N_\nHI,z)\simeq 10^{-12.1}\,
N_{\nHI,13}^{-1.5}\,(1+z)^{2.5}$ 
is the bivariate distribution of \HI columns and redshifts (cf. Kim \etal 
1997), and $cdt/dz$ is the line element in a Friedmann cosmology. 
From these relations one finds a mass fraction $f_m\simeq 0.6\,h$ which is 
independent of \HI column, i.e. most of the baryons at this epoch lie in the 
range $12.5<\log N_\nHI<14.5$ and are distributed equally per decade in column 
density (Zhang \etal 98). The volume filling factor of \Lya forest clouds is 
\beq
f_v=f_m\,{\overline\rho_b \over \rho_b}\approx 0.75\,h\,N_{\nHI,13}^{-0.7}. 
\eeq
Hence, in this picture, the metals associated with $\log N_\nHI\lta 14.2$
filaments fill a fraction $\gta 3\%$ ($h=0.5$) of intergalactic space, and are 
therefore far away from the high overdensity peaks where galaxies form, gas cools, 
and star formation takes place. Their chemical enrichment must then reflect more 
uniform (i.e. `early') rather than in--situ (i.e. `late') metal pollution.

\bigskip

\begin{figure}
\centerline{
\psfig{figure=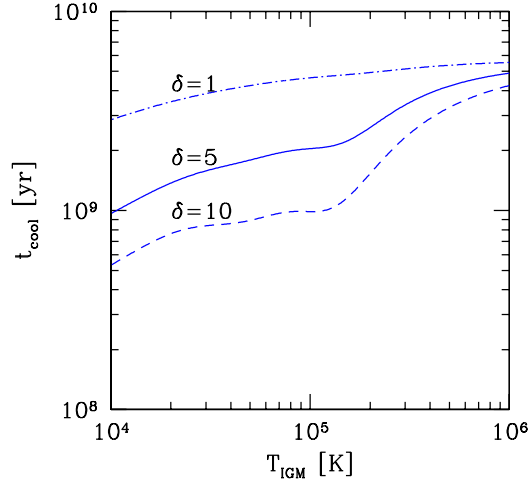,width=3.8in}}
\caption{Cooling timescale $t_{\rm 
cool}$ as a function of temperature $T_\igm$ for optically thin intergalactic 
gas at $z=3.5$. The medium is assumed to have an overdensity of $\delta=\rho_b/
\overline\rho_b=1$ ({\it dash-dotted line}),  5 ({\it solid line}), and 10 
({\it dashed line}), primordial abundances, and to be irradiated by a 
quasar-dominated UV/X-ray background.} 
\label{fig2} 
\end{figure}

\ni (2) At $z=3-3.5$, clouds with $N_\nHI\gta 10^{14.7}\,\cmm$ show a spread 
of at most an order of magnitude in their metallicity, and their narrow line 
widths require that they be photoionized and cold rather than collisionally 
ionized and hot (Songaila \& Cowie 1996). At these redshifts, hot rarefied gas,
exposed to a metagalactic ionizing flux, will not be able to radiatively cool 
within a Hubble time. The cooling time can be defined as the ratio of the 
specific energy content to the radiative cooling rate,
\beq
t_{\rm cool}={1.5nkT\over n_\nH^2\,\Lambda},
\eeq
where $k$ is Boltzmann's constant, $T$ is the gas temperature, and 
$\Lambda$ is the radiative cooling 
function. Here we have computed $\Lambda$ for a primordial  
plasma\footnote{At the low metallicities (less than 1\% solar) typical of \Lya 
forest clouds, the thermal behavior can be modeled to a good approximation by a 
gas with primordial abundances (e.g. Sutherland \& Dopita 1993).}\, with helium 
fraction by mass equal to 0.25, and total number density of all species $n$.
The cooling function, based on the
rates given in Hui \& Gnedin (1997), was calculated in the presence of a 
quasar--dominated UV/X--ray background (Haardt \& Madau 1996; Madau \& 
Efstathiou 1999); in this case the medium
is highly photoionized, collisions are not important in determining the
ionization structure (although we include them self--consistently), and 
collisional excitation cooling of \HI and \HeII becomes ineffective 
(Efstathiou 1992). 
The cooling timescale for optically thin intergalactic gas at 
different overdensities ($\delta\equiv \rho_b/\overline\rho_b=1, 5,$ and 10) 
and redshift $z=3.5$ is shown in Figure \ref{fig2}: it is longer than or at 
best comparable to the expansion timescale,
\beq
{1\over H}={1\over H_0 (1+z)^{3/2}}=10^9\,h^{-1}\left({1+z\over 
4.5}\right)^{-3/2}\; {\rm yr},
\eeq
at all temperatures.

While it is possible that some metals were dispersed 
in intergalactic space at late times, as hot pressurized bubbles of shocked 
wind and SN ejecta escaped the grasp of massive galaxy halos and expanded, cooling 
adiabatically, into the surrounding medium, such a delayed epoch of galactic 
super--winds would have severely perturbed the IGM (since the kinetic energy of 
the ejecta is absorbed by intergalactic gas), raising it to a higher
adiabat and producing variations of the baryons relative to the dark matter: 
\Lya forest clouds would not then 
be expected to closely reflect gravitationally induced density 
fluctuations in the dark matter distribution, and the success of hydrodynamical
simulations in matching the overall observed properties of \lya absorption 
systems would have to be largely coincidental.\footnote{Assume, for example, 
that the 
chemical enrichment of intergalactic gas was due to the numerous population 
of Lyman--break galaxies (LBGs) observed at $z=3$. With a comoving space 
density above $m_\star+1=25.5$ of $0.013\,h^3\,$Mpc$^{-3}$ 
(Steidel \etal 1999), a 1\% filling factor would be obtained if      
each LBG produced a metal--enriched bubble of proper radius equal to 
about $140\,h^{-1}$kpc. To fill such a bubble in $5\times 10^8\,$yr, the ejecta 
would have to travel at an average speed close to $600\,\kms$ (for $h=0.5$),
with characteristic postshock temperatures in excess of 2 million degrees.}
In contrast, the observed narrow Doppler widths could be explained if the 
ejection of heavy elements at velocities exceeding the small escape speed of 
subgalactic systems were to take place at very high redshifts. Hot enriched 
material cools more efficiently at these early epochs, since 
$Ht_{\rm cool}\propto (1+z)^{-3/2}$ and the Compton cooling time of the 
shocked ionized ejecta off cosmic microwave background (CMB) photons, 
\beq
t_{\rm comp}={3m_ec\over 4 \sigma_T aT_{\rm CMB}^4}=2.3\times 10^8 \left({1+z
\over 10}\right)^{-4}\; {\rm yr},
\eeq
is shorter than the expansion timescale.
Here $m_e$ is the electron mass, $\sigma_T$ the Thomson cross--section, $a$
the radiation constant, and $T_{\rm CMB}=2.725\,(1+z)\,$K the temperature of 
the CMB (Mather \etal 1999). Pregalactic outflows will propagate with typical
velocities of a few tens of $\kms$ into a dense IGM 
which has been pre--photoionized by the same massive stars which later explode 
as SNe, and the expansion of the metal--enriched bubbles will be halted by the 
external pressure. By $z=3$ any residual peculiar velocity would have been
redshifted away by a factor of 2--3, the \Lya forest would be hydrodynamically 
`cold', and the intergalactic baryons would have relaxed again under the 
influence of dark matter gravity.  

\bigskip

\ni (3) In a CDM universe, structure formation is a hierarchical process in 
which non linear, massive structures grow via the merger of smaller initial
units. Large numbers of low--mass galaxy halos are expected to form at 
early times in these popular cosmogonies, perhaps leading to an era of 
widespread pre--enrichment and preheating. The Press--Schechter (1974, hereafter
PS) theory for the evolving mass function of dark matter halos predicts a 
power--law dependence, $dN/d\ln m\propto m^{(n_{\rm eff}-3)/6}$, where $n_{\rm 
eff}$ is the effective slope of the CDM power spectrum, $n_{\rm eff}\approx -2.5$ 
on subgalactic scales. As hot, metal--enriched gas from SN--driven winds escapes 
its host halo, shocks the IGM, and eventually forms a blast wave, it sweeps a 
region of intergalactic space which increases with the $3/5$ power of the 
energy $E$ injected into the IGM (in the adiabatic Sedov--Taylor phase).    
The total fractional volume or porosity, $Q$, filled by these `metal bubbles' 
per unit explosive energy density $E\,dN/d\ln m$ is then 
\beq
Q\propto E^{3/5}\,dN/d\ln m\propto (dN/d\ln m)^{2/5}\propto m^{-11/30}.
\eeq
Within this simple scenario it is the star--forming objects with the smallest 
masses which will arguably be the most efficient pollutant of the IGM on 
large scales. Note, however, that since the cooling time of collisionally ionized 
high density gas in small halos at high redshifts is much shorter than the 
then Hubble time, virtually
all baryons are predicted to sink to the centers of these halos in the absence
of any countervailing effect (White \& Rees 1978). Efficient feedback is then 
necessary in hierarchical clustering scenarios to avoid this `cooling 
catastrophe',
i.e. to prevent too many baryons from turning into stars as soon as the first 
levels of the hierarchy collapse. The required reduction of the stellar 
birthrate in halos with low circular velocities may result from 
the heating and expulsion of material due to OB stellar winds and repeated 
SN explosions from a burst of star formation. 

\section{Basic theory}

\subsection{Dark matter halos}

To model the structural properties of subgalactic systems we will 
neglect the gravitational potential due to visible mass, and assume that
virialized dark matter halos, formed through hierarchical 
clustering, have a universal (spherically averaged) NFW density profile, 
\beq
\rho(r)={\rho_{\rm crit}\,\delta_c\over cx (1+cx)^2}, \label{eq:NFW}
\eeq
where $x\equiv r/r_\vir$, $r_\vir$ is the `virial' radius of the system, i.e.
the radius of the sphere encompassing a mean overdensity of 200, $c$ is the 
halo concentration parameter, $\delta_c=(200/3)c^3/F(c)$ is a characteristic 
overdensity, and
\beq
F(t)\equiv \ln(1+t)-{t\over 1+t}.
\eeq
The mass of the halo within the virial radius is $M=(4\pi/3)200\rho_{\rm 
crit}r_\vir^3$. Equation (\ref{eq:NFW}) implies a circular velocity, 
\beq
v_c^2(r)={GM(r)\over r}=V_c^2\,{F(cx)\over xF(c)},  \label{eq:vc}
\eeq
where $V_c^2\equiv GM/r_\vir$. 
Gas at radius $r$ will escape from the gravitational 
potential well only if it has a velocity greater than 
\beq
v_e^2(r)=2\int_r^\infty {GM(r')\over r'^2}dr'=2V_c^2\,{F(cx)+{cx\over 1+cx}
\over xF(c)}. 
\label{eq:ve}
\eeq
The escape speed is maximum at the center of the halo, $v_e^2(0)=2V_c^2c/
F(c)$. 

To proceed further, we follow the algorithm described in the appendix of 
NFW and compute the concentration parameter 
(or, equivalently, the characteristic density contrast $\delta_c$) 
of dark matter halos as a function of their mass in a SCDM 
model with $\Omega_M=1$, $h=0.5$, $\sigma_{8}=0.63$, and $n=1$.
The algorithm assigns to each halo of 
mass $M$ identified at redshift $z$ a collapse redshift $z_{\rm coll}$, 
defined 
as the time at which half of the mass of the halo was first contained in 
progenitors more massive than some fraction of the final mass.   
The assumption that the characteristic density of a halo is 
proportional to the critical density at the corresponding $z_{\rm coll}$ implies
\beq
\delta_c(M,z)\propto \left({1+z_{\rm coll}\over 1+z}\right)^3.
\eeq
Since lower mass systems generally collapse at higher
redshift, when the mean density of the universe is higher, at any given time 
low--mass halos will be more centrally concentrated than high--mass ones. For 
$M=10^8\,h^{-1}\,\msun$ and $z=9$, one finds $(z_{\rm coll}, c)=(12.2, 4.8)$. At 
the same redshift, a $10^{9}\,\msun$ ($10^7\,\msun$) halo would have
$(z_{\rm coll}, c)=(11.9, 4.7)$ [$(z_{\rm coll}, c)=(12.5, 4.9)$].

To study in detail the impact on the IGM of an episode of pregalactic star 
formation at $1+z\lta 10$, we will assume in the following a `typical' 
concentration parameter of $c=4.8$. At these epochs, the dark matter
halo of a subgalactic system will be characterized by a virial radius
\beq
r_\vir=0.76\,{\rm kpc}~M_8^{1/3}\, h^{-1}\,\left({1+z\over 10}\right)^{-1},
\eeq
a circular velocity at $r_\vir$ 
\beq
V_c=24\,{\kms}\,M_8^{1/3}\,\left({1+z\over 10}\right)^{1/2},
\eeq
and a virial temperature
\beq
T_\vir={GM\over r_\vir}{\mu m_p\over 2k}=10^{4.5}\,{\rm K}~ M_8^{2/3}\,\mu
\left({1+z\over 10}\right),
\eeq 
where $\mu$ is the mean molecular weight ($\mu=0.59$ for a fully ionized 
hydrogen/helium gas) and $M_8$ is the halo mass in units of $10^8 h^{-1}\,\msun$. 
The escape speed at the center is
\beq
v_e(0)=77\,{\kms}~ M_8^{1/3}\,\left({1+z\over 10}\right)^{1/2}.
\eeq
Note that high--resolution 
N--body simulations by Bullock \etal (2001) indicates that high--redshift halos
are actually less concentrated that expected from the NFW prediction.
In this case we may be slightly overestimating the escape speed from 
subgalactic systems.

\subsection{Halo gas cooling}

If gas collapses and virializes along with the dark matter perturbation to an 
isothermal distribution, it will be shock heated to the virial temperature and
settle down to a density profile 
\beq
\ln\rho_{\rm gas}(r)=\ln\rho_0\,-{\mu m_p\over 2kT_\vir}[v_e^2(0)-
v_e^2(r)]
\eeq
(Makino, Sasaki, \& Suto 1998). 
The central gas density $\rho_0$ is determined by the condition that the 
total baryonic mass fraction within the virial radius is equal to 
$\Omega_b$ initially: 
\beq
{\rho_0\over \rho_{\rm crit}}={{200\over 3}c^3\Omega_be^A
\over \int_0^c (1+t)^{A/t}\,t^2 dt}=840\,h^{-2},
\eeq
where $A\equiv 2c/F(c)$. At the virial radius $\rho_{\rm gas}(r_\vir)=0.00144
\rho_0$. 

\begin{figure}
\centerline{
\psfig{figure=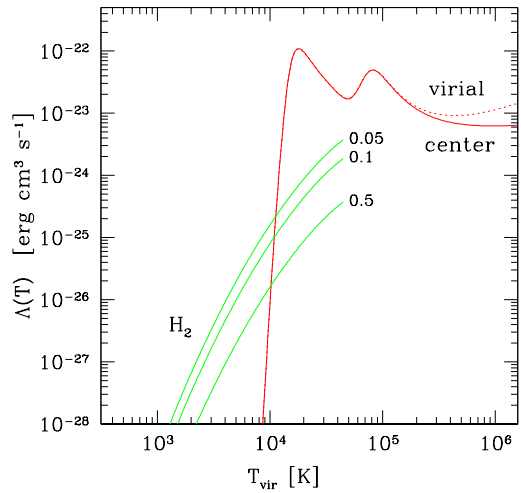,width=3.8in}}
\caption{Equilibrium cooling rate at the center ({\it solid lines}) and 
virial radius ({\it dashed line}) of an isothermal
halo at $z=9$, as a function of virial temperature $T_\vir$,
and in the absence of a photoionizing background (i.e. prior to the reionization
epoch). The halo has an assumed baryonic mass fraction of $\Omega_b=0.019h^{-2}$. 
The gas density dependence of the cooling function at high temperatures is due 
to Compton cooling off comic microwave background photons. The labeled curves 
extending to low temperatures show the contribution due to H$_2$ for three
assumed values of the metagalactic flux in the Lyman-Werner bands, $4\pi J_{\rm 
LW}$ (in units of $10^{-21}\,\units$).
\label{fig3}}
\end{figure}

\begin{figure}
\centerline{
\psfig{figure=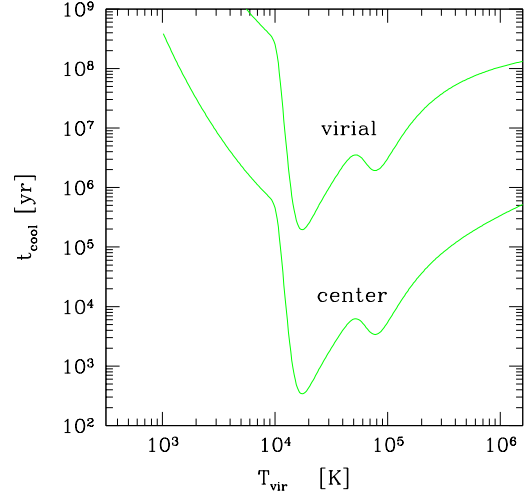,width=3.8in}}
\caption{Equilibrium cooling time  
at the center and virial radius of an isothermal
halo at $z=9$, as a function of virial temperature $T_\vir$,
and in the absence of a photoionizing background. Assumptions as in Fig. 
\ref{fig3}, with $4\pi J_{\rm LW}=0.5 \times 10^{-21}\,\units$. 
\label{fig4}}
\end{figure}

Figures \ref{fig3}--\ref{fig4} show the cooling function $\Lambda$ and timescale
$t_{\rm cool}$ at the center and virial radius of an isothermal halo 
at $z=9$, as a function of the halo virial temperature.
Cooling by \Lya line radiation becomes inefficient for gas 
temperatures below $2\times 10^4$~K: in a gas of primordial
composition at such low temperatures the main coolant is radiative 
de--excitation of the rotational and vibrational states of molecular hydrogen. 
Objects relying on H$_2$ cooling are usually referred to as Population III 
objects. Primordial H$_2$ is produced with a fractional abundance of $f_{\rm 
H_2} \approx 2\times 10^{-6}$ at redshifts $\lta 110$ via the H$^-$ formation 
channel. 
Starting from this low value the H$_2$ abundance increases in collapsing
pregalactic clouds, molecular cooling becomes efficient, and stars can form.
While most of the Lyman--continuum photons produced by these stars are quickly 
absorbed by the dense \HI disk layers, radiation in the Lyman--Werner bands 
escapes 
into the IGM to form a soft UV cosmic background which can photodissociate 
H$_2$ via the Solomon process, thus inhibiting 
further star formation. An estimate of the H$_2$ equilibrium fraction under
these conditions can be obtained by balancing the rates for the above 
processes: 
\beq
f_{\rm H_{2}}={k_f n_e \over k_d}= 4.8\times 10^{-9}\,T_\vir^{0.88}
(4\pi J_{\rm LW,21})^{-1},
\eeq
where $J_{\rm LW,21}$ is the specific metagalactic flux in the 11.2--13.6 eV 
energy range in units of $10^{-21}\,\uvunits$, 
$n_e$ is the residual electron density after recombination,
and we have taken the formation ($k_f$) and dissociation
($k_d$) rates of Abel \etal (1997). Using the above relation, we have 
included the contribution to the cooling rate per particle due to molecular 
hydrogen (Martin, Schwarz, \& Mundy 1996) in halos of different virial 
temperature. The derived cooling function and cooling timescale at $z=9$ are 
depicted in Figures \ref{fig3}-\ref{fig4} for different values of 
$J_{\rm LW,21}$. Note that here and below we are implicitly assuming that the 
universe is reionized at a redshift $z<9$: this is because,
after reionization occurs, there will be a universal background of photons
above 13.6 eV which inhibits the formation of dwarf galaxies both by reducing 
the cooling rate of gas within halos with $T_\vir\lta 5\times 10^4\,$K, and 
also by suppressing the accretion of high entropy ionized gas into 
subgalactic fragments (e.g. Efstathiou 1992; Thoul \& Weinberg 1996; 
Gnedin 2000). 

\begin{figure}
\centerline{
\psfig{figure=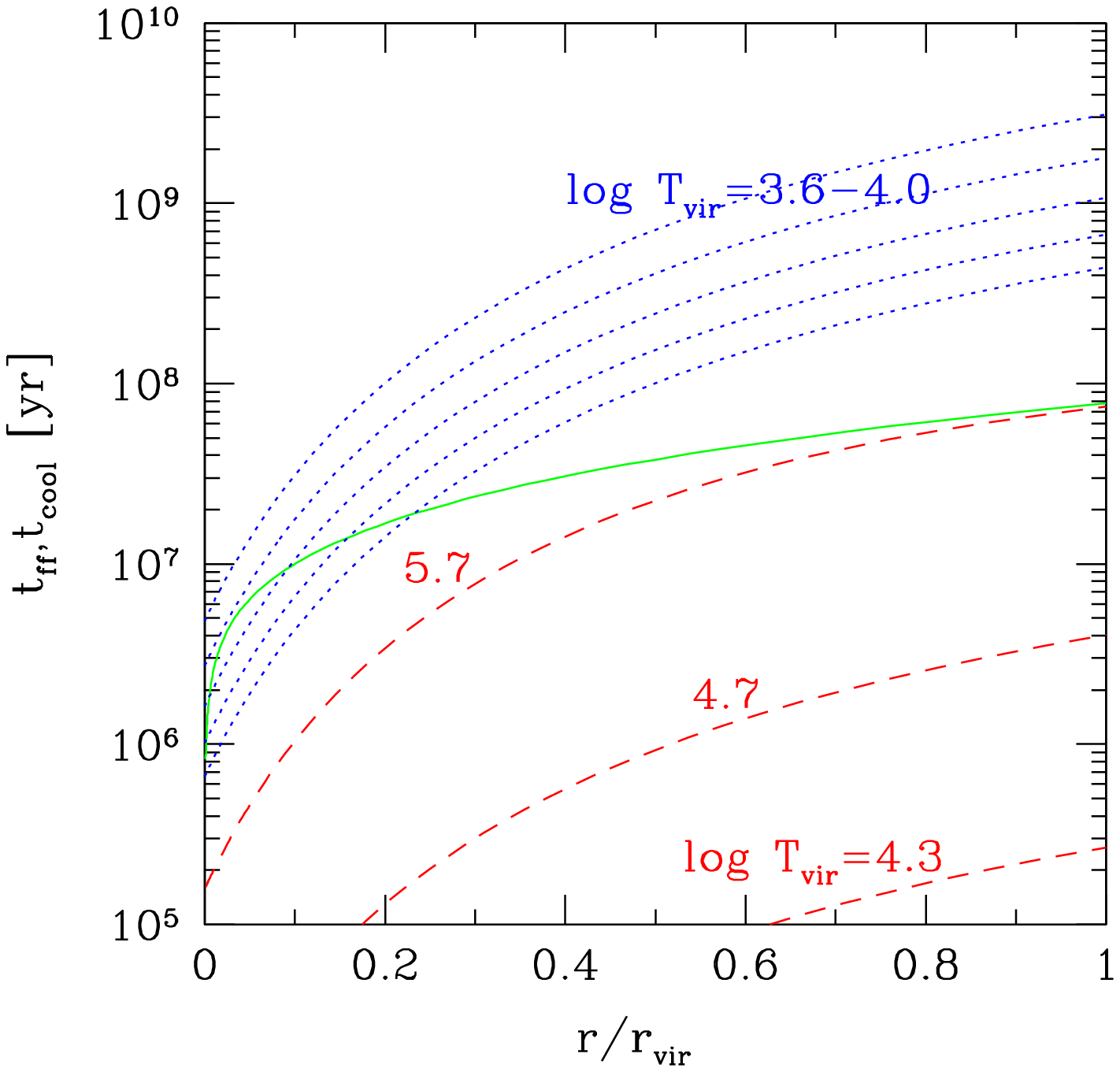,width=3.8in}}
\caption{Halo cooling times at $z=9$ as a function of radius for 
$\log T_\vir=4.3, 4.7, 5.7$ ({\it dashed curves}). The gas is assumed to be
isothermal and in collisional ionization equilibrium. {\it Dotted 
curves:} same for halos with virial temperatures in the
range $3.6\le \log T_\vir \le 4.0$, where cooling is dominated by H$_2$ ($4\pi 
J_{\rm LW}=0.5 \times 10^{-21}\,\units$ case). 
{\it Solid curve:} gravitational free--fall time of a NFW halo at $z=9$ 
($h=0.5$). 
\label{fig5}}
\end{figure}

\begin{figure}
\centerline{
\psfig{figure=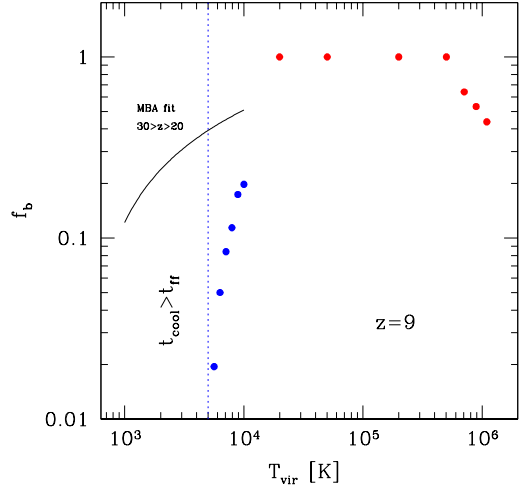,width=3.8in}}
\caption{Fraction $f_b$ of the total halo gas mass at $z=9$ that cools 
faster than the local free--fall time, i.e. never reaches hydrostatic 
equilibrium. In halos with $4.3<\log T_{\rm vir}<5.7$ $f_b=1$, i.e. all the 
accreted gas can cool. For comparison, the fit for $f_b$ derived
by Machacek \etal (2000, MBA) from their hydrodynamics simulations is  
also shown.
These numerical results have 
been obtained in the redshift range $20\lta z\lta 30$, where because of the 
higher mean baryon density the rate of formation of H$_2$ molecules 
is much faster than at the 
redshift of 9 considered here. 
Some significant differences also arise from the different density and 
temperature profiles of Machacek \etal simulated halos.
The gas in systems with $T_\vir<5000\,$K ({\it dotted vertical line}) cannot
cool faster than the local free--fall time at any radius.
\label{fig6}} 
\end{figure}

The fraction of the baryonic content of a halo that can actually cool and reach
the center is determined by the balance between the cooling and the dynamical
timescales of the systems.
As shown in Figure \ref{fig5}, for $4.3<\log T_\vir<5.7$ rapid cooling by 
atomic hydrogen and ionized helium can occur at these epochs on timescales 
much shorter than the free--fall time,
$$
t_{\rm ff}(r)=\int_0^{r} {dr'\over \sqrt{v_e^2(r')-v_e^2(r)}}=
2.2\times 10^7\,h^{-1}\,{\rm yr}
$$
\beq
~~~~\times \left({1+z\over 10}\right)^{-3/2}\,
\int_0^x dx'\,[{\cal F}(cx')-{\cal F}(cx)]^{-1/2}, 
\eeq 
for a gas element at all radii $r<r_\vir$,
where ${\cal F}(cx)\equiv [F(cx)+cx/(1+cx)]/[xF(c)]$. Therefore, for masses
in the range $10^8\,h^{-1}\lta M\lta 10^{10}\,h^{-1}\,\msun$, infalling gas
never comes to hydrostatic equilibrium, but collapses to the center at the 
free--fall rate. 
Outside this mass range, i.e. when cooling is dominated by H$_2$ (at the 
low--end) and free--free emission (at the high--end), the halo gas can be 
pressure--supported and form a quasi--static hot atmosphere. If we denote with 
$r_{\rm cool}$ the radius where the cooling time is equal to the 
free--fall time, a parameter $f_b$ can now be defined 
as the ratio between the gas mass within $r_{\rm cool}$ 
and the total baryonic mass within the virial radius, $\Omega_bM$.
This gas fraction is plotted in Figure \ref{fig6} as a function of virial 
temperature. In halos with $f_b=1$ all the accreted gas can cool immediately, 
and the 
supply of cold gas for star formation is only limited by the infall rate.
Conversely, in systems with $f_b\ll 1$, the supply of cold gas is regulated 
by the longer cooling timescale everywhere but for a small amount of gas in the
very central region of the halo. When weighted with the steep PS mass function, 
it is the gas at the peak of the cooling curve -- i.e. gas in subgalactic 
systems with masses comparable to the masses of present--day dwarf ellipticals -- 
than may be more readily available to be 
transformed into stars on short timescales, and give origin to explosive 
multi--SN events.   

In the adopted cosmology (SCDM with $h=0.5$ and rms mass fluctuation 
normalized at present to $\sigma_8=0.63$ on spheres of $8\,h^{-1}\,$Mpc),
$M=10^8\,h^{-1}\,\msun$ halos would be collapsing at $z=9$
from 2--$\sigma$ fluctuations. At this epoch, more massive halos with 
$M=10^{10}\,h^{-1}\,\msun$, while able to cool rapidly, would be collapsing from 
3--$\sigma$ peaks and be too rare to produce significant amounts of 
heavy elements (in a gaussian theory the peaks above 3--$\sigma$ 
contain $\sim 5\%$ as much mass as those above 2--$\sigma$ at a given epoch), 
unless they were able somehow to form stars 
more efficiently than lower--mass objects. Even if this were the case, however,
the SN ejecta would escape the grasp of these more massive halos with less ease,
and the fractional volume of the IGM filled by their metal bubbles
be correspondingly small. Halos from 1--$\sigma$ fluctuations would be more 
numerous and contain most of the 
mass, but with virial temperatures of only a few hundred degrees they would 
likely be unable to cool ($f_b\ll 1$) via H$_2$ before reionization actually 
occurs. 

A detailed history of the chemical enrichment of the IGM should 
include the contribution to its metal content from all levels of the mass 
hierarchy at every epoch and is beyond the scope of this paper. 
Recent computations (Haiman \etal 2000; Ciardi \etal 2000; Ricotti, 
Gnedin \& Shull 2001) trying to assess the importance of radiative feedback 
on Pop III halos have reached different conclusions. Although it is clear that 
a soft UV 
radiation field tends to prevent the cooling and collapse of these objects, and that
star formation inside Pop III systems shuts off well before reionization,
quantifying this effect is difficult as the field intensity depends on poorly
unknown parameters as the star formation efficiency and the escape fraction of 
ionizing photons from the star formation sites into the IGM. Additional
complications are introduced by locally produced Lyman--Werner photons which 
photodissociate H$_2$ molecules in the parent halo gas (Omukai \& Nishi 1999) and 
by a hard radiation components which could boost the formation rate of 
molecular hydrogen (Haiman \etal 2000) by increasing the electron fraction and 
thus feeding the 
H$^-$ formation channel. Given these uncertainties we will take here a
conservative approach and assume that all objects below the critical mass 
for \Lya efficient cooling are suppressed. Below we will 
focus on the role played by what we have suggested might be the most efficient
pollutant of the IGM on large scales, subgalactic systems with masses 
$\sim 10^8\,h^{-1}\,\msun$ at redshift 9, when large numbers of them grow
non--linear and collapse. 

\subsection{Energy injection by SNe}

When gas cools well below its initial virial temperature and infalls 
to the center, it fragments into clouds and then into stars.
Unless the stellar IMF is dramatically bottom--heavy, stars more massive
than $8\,\msun$ will form in the inner densest (self--gravitating) regions of 
the halo, eventually
releasing their binding energy in a supernova explosion, returning most of 
the metals to the ISM, and injecting about $10^{51}\,$erg per event in kinetic 
energy. 
We shall assume in the following that (i) all SN explosions take place at 
the center of the halo, and (ii) our stellar population forms in a time 
interval which is short compared to the lifetime of massive stars.
These two simplifying assumptions are delicate and need to be discussed.
The first hypothesis implies spatial coherency
among explosions, which in turn ensures that all the energy released 
by different SNe is used to drive the same `superbubble', producing
a cumulative, multi--SN event. This condition is not generally met in
large galaxies like our own, where superbubbles occur essentially at random in 
the disk. For objects with small circular velocities, however, the spatial 
coherence is essentially guaranteed by the fact that the size of the region 
where star formation occurs is comparable or smaller than the characteristic 
scale of the bubbles. Indeed, if a rotationally--supported exponential disk 
with scale length $r_d$ forms in subgalactic fragments, and the specific 
angular momentum of the disk material is the same as that of the halo, then 
angular momentum conservation fixes the collapse factor to $r_\vir/r_d=\sqrt{2}/
\lambda$, where $\lambda$ is the spin 
parameter of the halo ($\approx 0.05$ from N--body simulations, Barnes \& 
Efstathiou 1987), and the equality applies to dark halos treated
as singular isothermal spheres. Our fiducial $M=10^8\,h^{-1}\,\msun$ system 
at $z=9$ has $r_\vir=0.76\,h^{-1}\,$kpc and $r_d=27\,h^{-1}\,$pc. 
If we assume, for simplicity, that the self--gravitating disk of mass $M_d$   
follows an isothermal vertical profile with a thermal speed $c_s=10\,\kms$,
typical of gas which is continuosly photoheated by stars embedded within the
disk itself, then its scale height at radius $r_d$ is $h/r_d=\sqrt{16e}\,
\lambda (M/M_d) (c_s/V_c)^2$, and the disk is rather 
thick (e.g. Wood \& Loeb 2000). Within a scale height the gas density 
is approximately constant,
and the radius $R_s$ of a supernova remnant in a uniform medium of density 
$n$ and in the pressure--driven `snowplough' stage is given by 
$R_s=(E_{51}/n)^{1/7}\,t_{\rm yr}^{2/7}$ pc (McKee \& Ostriker 1977). It will
take then only 
$10^5$ yr for an individual bubble to grow bigger than a disk scale length.

The second assumption  has to do with the ability of
the collapsing system to transform the cold material into stars on timescales 
shorter than a few $\times 10^7$~yr, the lifetime of a typical SN progenitor. 
A larger spread in the stellar birth times is likely to decrease the final 
number of SNe, as ionizing photons from the first massive stars will reheat
and ionize the infalling gas, thus inhibiting the formation of
subsequent stars. It is difficult to assess the validity of such hypothesis, 
given our current poor understanding of star formation. 
While in a typical OB association in the Milky Way roughly three SN per million 
years will occur (Heiles 1990), a much higher rate may be sustained in 
pregalactic systems due to the very short cooling timescale of the halo central 
regions.
If we accept these two assumptions, which can only be met by 
subgalactic fragments, we can calculate the effects of energy deposition 
in a star--forming halo. Usually the distinction is made between blowout,
a partial removal of the gas from a galaxy, and blow--away, in which 
the entire gas content is ejected back into the IGM (e.g. MacLow \& Ferrara 
1999). For a spherical system, as our admittedly idealized collapsing halo, the 
two terms are synonymous. Therefore, the superbubble will escape into the IGM 
only if it can lift out of the halo its entire gas mass. 
The amount of material transformed into stars can be parameterized as
\begin{equation}       
M_\star = \Omega_b f_b f_\star M. 
\end{equation}       
The cooled fraction of baryons, $f_b$, has already been discussed above and found
equal to unity in such halo mass range. There are no firm estimates for the star 
formation efficiency $f_\star$, however, and we consider it as a free parameter 
of our model. We will analyze three limiting scenarios: a low
($f_\star=1$\%), a medium ($f_\star=10$\%), and a high ($f_\star=50$\%) 
efficiency case. For $h^2\Omega_b=0.019$ and $h=0.5$, 
$M_\star$ ranges between 1.5 and 75$\times 10^5\,\msun$. 
With a comoving space density of dark halos with masses above $10^8\,h^{-1}\,\msun$ 
of about $80\,h^3\,$Mpc$^{-3}$ at $z=9$ (see \S\,5) this corresponds to a 
stellar density parameter of $0.002\,f_\star$, i.e. between 0.4\% and 20\%  
of the total stellar mass inferred at the present epoch (Fukugita, Hogan,
\& Peebles 1998). We note that, on this assumption, $f_\star$, and hence the 
early luminosity of these systems, would exceed that which is predicted by the usual 
low-mass extrapolation of  CDM galaxy formation models (cf. White \& Frenk 1991). In 
these scenarios $f_\star$ is postulated to decline steeply in 
shallow potential wells, thereby reducing the population of low--luminosity 
galaxies and avoiding the so--called `cooling catastrophe'.
We also stress 
that the value $f_\star=50$\% is only meant to represent an extreme case
and should really be considered as 
an upper limit to the star formation efficiency of our fiducial system. This is 
because, when most mechanical energy is injected by SNe after $3\times 
10^7\,$yr (the main sequence lifetime of a $8\,\msun$ star) and 
SN--driven bubbles propagate into the galaxy halo quenching further star
formation, only material within $0.4\,r_\vir$ has actually had time to cool, 
free--fall to the center, and form stars (Fig. \ref{fig5}). With the adopted density 
profile, the gas within the radius where $t_{\rm ff}=3\times 10^7\,$yr makes 
about 40\% of the total baryonic 
mass $\Omega_bM$ of the halo. In general, one would expect significantly lower 
efficiencies than $50\%$ as the conversion of cold gas into stars is limited by 
the increasing fractional volume occupied by supernova remnants.
Additional constraints on $f_\star$ may come from the observed metallicity
distribution of the most metal poor stars in the Milky Way halo, although
the results will be subject to uncertainties in the low--mass end of the 
IMF.  

The IMF also determines the number of supernova events.
Conservatively, we assume here a Salpeter IMF, with upper 
and lower mass cut--offs equal to $M_u=120\,\msun$ and $M_l$, respectively. 
The value of $M_l$ is varied from the standard  case $M_l=0.1\,\msun$
to a value appropriate for a top--heavy IMF often predicted       
for very low metallicity stars, $M_l=5\,\msun$, and up to the
extreme case $M_l=30\,\msun$, in which every star formed explodes as a supernova 
and many of them may eventually end their life as a black hole.

\subsection{Mechanical luminosity}

The halos under study produce a rather limited amount of very massive stars, 
and the IMF is then sampled in a stochastic manner. To 
determine the time--dependent mechanical luminosity injected by SN explosions 
when a fraction $f_\star$ of the gas mass is converted into stars, we have 
repeatedly sampled the IMF by using  a Monte Carlo method until the
desired total stellar mass was reached. This procedure yields for 
$M_l=0.1\,\msun$
and $f_\star= (1\%, 10\%, 50\%)$ the following number of stars more massive
than $8\,\msun$ (hence of SNe): ${\cal N} = (1126, 11172, 55094)$, 
respectively. For this IMF, 1 SN occurs every $\nu^{-1}\approx 136\,\msun$
of stars formed, with an average stellar mass $\langle M_s\rangle =0.35\,\msun$. 
For $f_\star=1$\%, the choice $M_l=5\,(30)\,\msun$ gives 
${\cal N}=6076\,(3000)$, $\nu^{-1}=24\,(52)\,\msun$ and $\langle M_s\rangle
=13.5\,(52.7)\, \msun$. As we will see later, the differences in 
the efficiency and IMF will influence the fate and evolution of the ensuing 
superbubble and its metal content.

To determine the main sequence lifetime, $t_{\rm OB}(M_s)$, of massive stars
we have used a compilation of the data available in the literature (Schaller 
\etal 1992; Vacca, Garmany, \& Shull 1996; Schaerer \& de Koter 1997; Palla,
private communication), and derived the approximate fit
\beq
{t_{\rm OB}(M_s)\over {\rm Myr}}=\left\{
\begin{array}{rl} 33 \left({M_s\over 8\,\msun}\right)^{-3/2} 
\quad\mbox{$M_s \le 28.4\,\msun$},\\
3.4 \left({M_s\over 60\,\msun}\right)^{-1/2} \quad\mbox{$ M_s> 28.4\,\msun$}.
\end{array}\right. 
\eeq
The extrapolation to masses larger than $60\,\msun$ is quite uncertain; 
in general this massive stars are rare enough that this will not seriously affect 
our results.

We can now derive the mechanical luminosity of the massive--star association 
driving the superbubble. Since all stars are assumed to be born coevally 
in a single burst of star formation, the spread 
in the SN energy deposition is only due to the difference in $t_{\rm OB}$ for 
the various masses. The mechanical luminosity 
is defined as $L(t)=dE/dt$, where $E$ is the energy 
produced by the ensemble of SNe.  We further assume that each supernova 
releases $E_0=10^{51}\,$erg in kinetic energy. The derivation naturally 
accounts for the stochastic behavior of $L(t)$, which nevertheless has two 
clear    
features: {\it i)} a pronounced initial peak during the first 5 Myr after the
burst, caused by
the crowding of the explosions of the most massive stars which tend to 
have very similar ages (see expression for $t_{\rm OB}$ above); and 
ii) random oscillations around a mean value roughly equal to ${\cal N}E_0/{\rm 
max}\{t_{\rm OB}(M_s)\}$.  

\subsection{Superbubble evolution}

In this and the following sections we will model the evolution of SN--driven
bubbles as they blow out from our $10^8\,h^{-1}\msun$ fiducial halo, allowing
for radiative losses, gravity, external pressure, and thermal conduction.
Correlated multi--SN explosions will create large holes in the ISM of pregalactic
systems, enlarging pre--existing ones due to winds from their progenitors
stars. Most of the swept--up mass, both in the early adiabatic 
and in the following radiative phases, is concentrated in a dense shell
bounding the hot overpressurized interior, which yet contains enough mass
to thermalize the energy input of the SNe. 

Superbubbles are canonically studied by using the thin shell approximation
(Kompaneets 1960; Ostriker \& McKee 1988), which has been checked
against numerical simulations giving
excellent agreement (MacLow \& McCray 1988). The shell 
expansion, whose radius is denoted by $R_s$, is driven by the internal 
energy, $E_b$,  of the hot bubble gas. The pressure of such a gas (with adiabatic 
index $\gamma=5/3$) is therefore 
$P_b=E_b/2\pi R_s^3$. Hence, momentum and energy conservation yield the
relevant equations:
\begin{equation}
\label{tsh}        
{d\over dt}(V_s \rho \dot R_s) = 4\pi R_s^2 (P_b -P)- {GM(R_s)\over R_s^2}\rho V_s, 
\end{equation}
\begin{equation}
\label{tsh1}        
{dE_b\over dt}=L(t) -4\pi R_s^2P_b\dot R_s - V_s \overline n_{\nH,b}^2\Lambda 
(\overline T_b), 
\end{equation}
where the subscripts {\it s} and {\it b} indicate shell and bubble quantitites, 
respectively.
We have defined the volume enclosed by the shell as $V_s=(4\pi/3) R_s^3 $,
the dots represent time derivatives, and $\rho$ is the pressure of the 
ambient medium, taken to be equal to the halo gas 
density within $r_\vir$, and to the IGM background density at $z=9$ 
outside the virial radius. As at $r_\vir$ the halo is still about 60 times 
denser than the IGM, to avoid unphysical effects due to this jump we have 
allowed the two distributions to merge through an exponential transition of 
width $\Delta = 0.2\,r_\vir$. 
Finally, $\overline n_{\nH,b}^2\Lambda(\overline T_b)$ is the 
cooling rate per unit volume of the hot bubble gas, whose average hydrogen density 
and temperature are $\overline n_{\nH,b}$ and $\overline T_b$, respectively. The 
physical interpretation of the various terms is 
straightforward: in the momentum equation, the first term on the r.h.s. 
describes the  
momentum gained by the shell from the SN shocked wind, while the second term 
corresponds to the momentum lost to the local gravitational field. The terms
on the r.h.s. of the energy equation describe the mechanical energy input, 
the work done against the shell, and the energy losses due to radiation.
                                
It is important to remark here that we are actually neglecting the complicated 
web--like structure ubiquitous to three--dimensional cosmological hydrodynamical
simulations (e.g. Cen \etal 
1994; Zhang \etal 1995). In CDM universes, virialized systems form 
at the intersection of mildly overdense filaments, along which most of the mass 
accretion (inflow) actually occurs; outside the virial radius there 
will still be a power--law decrease in density within material 
that is flowing in, but of course the spherical assumption breaks down.
Moreover, prior to the reionization epoch, a significant fraction of all baryons 
are not distributed in a low--density IGM but actually condense into numerous 
small halos with virial temperatures 
above the cosmological Jeans mass. Such `minihalos' have not been yet resolved
in large--scale
cosmological simulations, and are expected to dominate the average gas clumping in 
the intergalactic medium (Haiman, Abel, \& Madau 2001).   
We have also ignored any effects due to possible inhomogeneities within the halo 
gas. The shallow slope of the CDM spectrum on mass scales $\lta 10^8\,\msun$    
leads to all small--scale fluctuations going nonlinear almost simultaneously 
in time. The evolution of these early halos will then be marked by frequent 
mergers, which could raise the gas to a higher adiabat by shock heating and
yield complex velocity and density fields within the `virial radius' (Abel
\etal 2000). In this first assessment of pregalactic outflows, we shall assume 
for simplicity that much of their structure and hydrodynamics can be understood 
from spherical profiles of the physical quantities.  

To determine the pressure $P$ of the ambient medium
we further assume that both the halo gas and the IGM are photoheated at a 
temperature of $10^4$~K by the SN progenitors. In general, the size of an 
intergalactic \HII region around a galaxy halo will depend on the H--ionizing
photon luminosity $\dot N_i$, on the fraction $f_{\rm esc}$
of these photons that can actually escape the dense star formation regions 
into the IGM, on the IGM mean density $\overline n_\nH$, and on the 
volume--averaged recombination timescale, $\overline t_{\rm rec}$. When 
the source lifetime $t_s$ is much less than $(\overline t_{\rm rec},H^{-1})$, 
however, as expected for a subgalactic halo shining for a few
$\times 10^7$ yr before being blown away by SN explosions (or in the case of 
a short--lived QSO; Madau \& Rees 2000), recombinations can be 
neglected and the evolution of the \HII region can be decoupled from the Hubble
expansion. The radius of the ionized zone is then
$$
R_I=\left({3 \dot N_i f_{\rm esc}t_s\over 4\pi\overline{n}_\nH}\right)^{1/3}
\approx (54~{\rm kpc})\,
\left({\Omega_b h^2 \over 0.02}\right)^{-1/3}
$$
\begin{equation}
\times \,\left({1+z\over 10}\right)^{-1}\,
\left({t_s\over 10^7\,{\rm yr}}\right)^{1/3}\,(\dot N_{52}f_{\rm esc})^{1/3},  
\label{ri}
\end{equation}
where $\dot N_i=10^{52}\,\dot N_{52}$~s$^{-1}$ is the 
ionizing photon luminosity due to $10^3$ massive stars distributed according to 
a Salpeter IMF.  $R_I$ will then be larger than the final size of the SN-driven 
superbubble (derived below) for values of the escape fraction greater than a 
few percent.

It is interesting to derive the timescale  at which the 
postshock gas enters the radiative phase. One can show (Weaver \etal 1977)
that when the ambient gas pressure, gravity, and cooling can be neglected, 
and $L(t) =$ const, the solution of the above equations is 
\begin{equation}
\label{anal} 
R_s = \left({125\over 154 \pi}\right)^{1/5} \left({L t^3\over 
\rho}\right)^{1/5}. 
\end{equation}
The temperature of the postshock gas is then
\begin{equation}
\label{tps} 
T_{\rm ps} = \left({3\mu m_p\over 16 k}\right) \dot R_s^2 = 4\times 10^4 
\left({L_{38} \over t^2_{\rm Myr} n}\right)^{2/5} {\rm ~K}.
\end{equation}
For $f_\star=1$\% and $M_l=0.1\,\msun$ one has $L_{38}\equiv L/(10^{38}$ erg s$^{-1})
\approx 10$; also, in the central region of our fiducial halo the total gas number 
density is $n\approx 10$~cm$^{-3}$. At such temperatures and densities the cooling 
rate per particle is $\approx 2\times 10^{-22}$~erg~s$^{-1}$. Hence the 
cooling time is $t_{\rm cool}(t=1\,{\rm Myr})\approx 1300$~yr, 
much shorter than the dynamical time 
of the system, ensuring that the shell forms very rapidly and justifying the use 
of the thin shell approximation.
These estimates also highlight the difference between a bubble produced by 
repeated SNe and a single point explosion. In the former case most of the 
energy of the bubble resides in the hot, very 
inefficiently radiating cavity gas generated by the continuous energy injection,
while most of the mass is contained in a thin layer that collapses to form 
the dense, cool shell.  

The final step consists in the determination of the cooling rate of
the bubble, which depends on the density and temperature of the hot 
interior. The thin--shell equations only provide a relation for
the product of these two quantities (\ie the pressure), so another
physical relation must be derived. Most of the bubble gas mass comes
from the conductive evaporation at the contact surface between the hot 
gas and the cold shell. The structure of conductive/cooling fronts 
has been studied by several authors (e.g. Cowie \& McKee 1977; McKee \& 
Begelman 1990; Ferrara \& Shchekinov 1993).
Under the conditions in which thermal conduction is unsaturated, the evaporative 
flow is steady, and radiative losses are negligible, the
rate at which gas is injected from the shell into the cavity is
\begin{equation}
\label{mevap}
{dM_{\rm ev}\over dt} = {16\pi\mu\eta\over 25 k} T_b^{5/2} R_s = C_1  T_b^{5/2} R_s,
\end{equation}
where $\eta=6\times 10^{-7}$ (we have assumed a Coulomb logarithm 
equal to 30) is the classical Spitzer thermal conduction coefficient. 

The interior distribution of temperature and density are known 
to obey a self--similar solution of the type $f(x)=f_c
(1-\zeta)^q$, where $q=2/5$ ($-2/5$) if $f$ represents the temperature (density);
$f_c$ is the value of the variable at the center and $\zeta\equiv r/R_s$, where
$r$ is the radial coordinate inside the bubble. According to 
this solution $P_b \propto n_b T_b =$ const, consistent with the hypothesis
implicitly made that SN shocks rapidly decay into sound waves inside $R_s$. 
By integrating the density profile up to $R_s$ we obtain the total mass in the 
bubble, $M_b= C_2 n_b R_s^3$, where $C_2=(125/39)\pi\mu m_p$. 
Differentiating this expression with respect to time and equating it to 
the evaporation rate (\ref{mevap}), we finally obtain an equation for the
evolution of the temperature $T_b$
\begin{equation}
\label{Tevol} 
{dT_b\over dt} = 3 {T_b\over R_s}\dot R_s + {T_b\over P_s} \dot P_s - 
 {23\over 10} {C_1\over C_2} {kT_b^{9/2}\over R_s^2 P_s}.
\end{equation}
This relation closes the system equations (\ref{tsh})--(\ref{tsh1}), and 
allows us to derive the temperature and density structure of the 
bubble and its cooling rate. 

\section{Numerical results}       

\begin{figure}
\plottwo{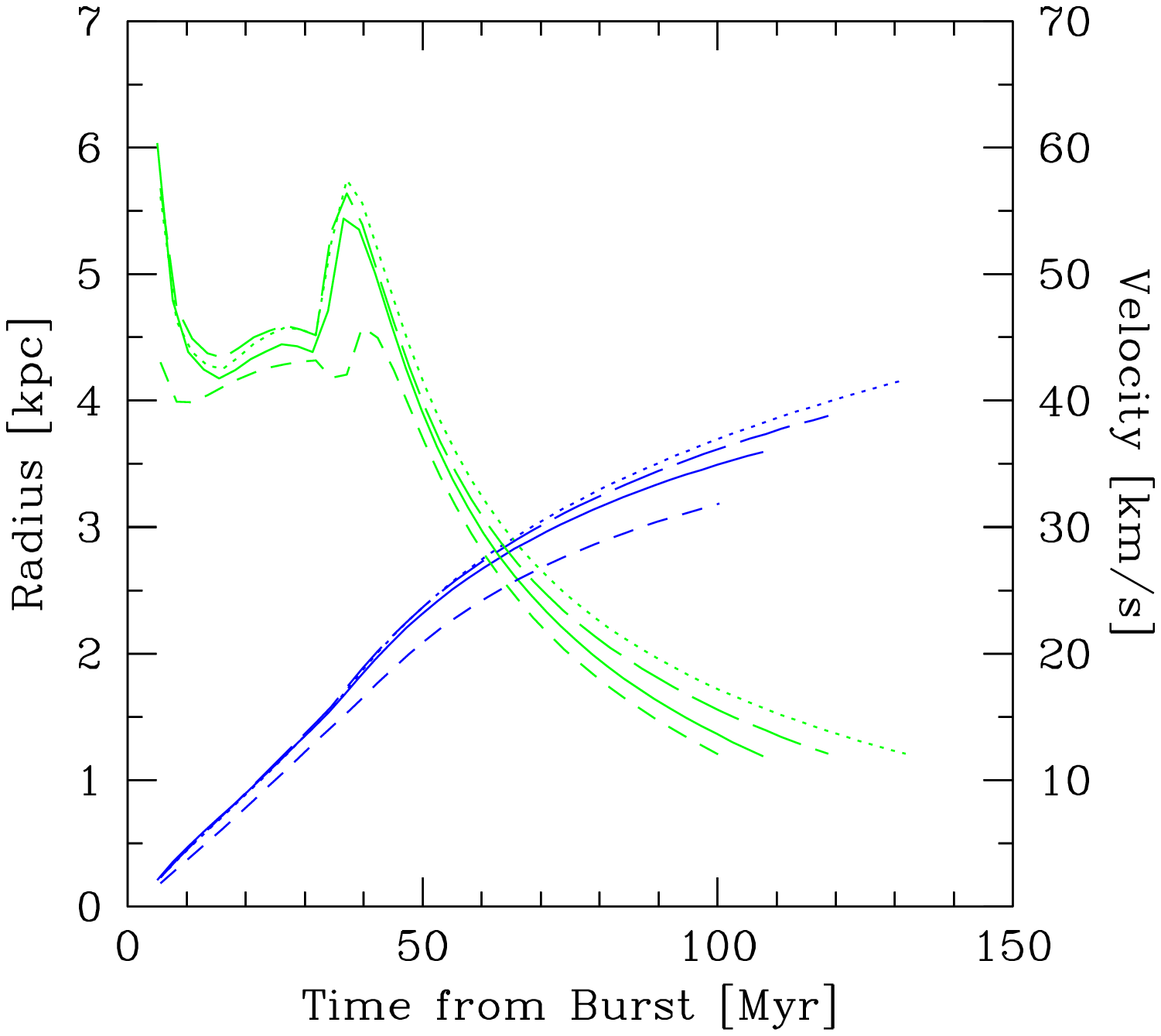}{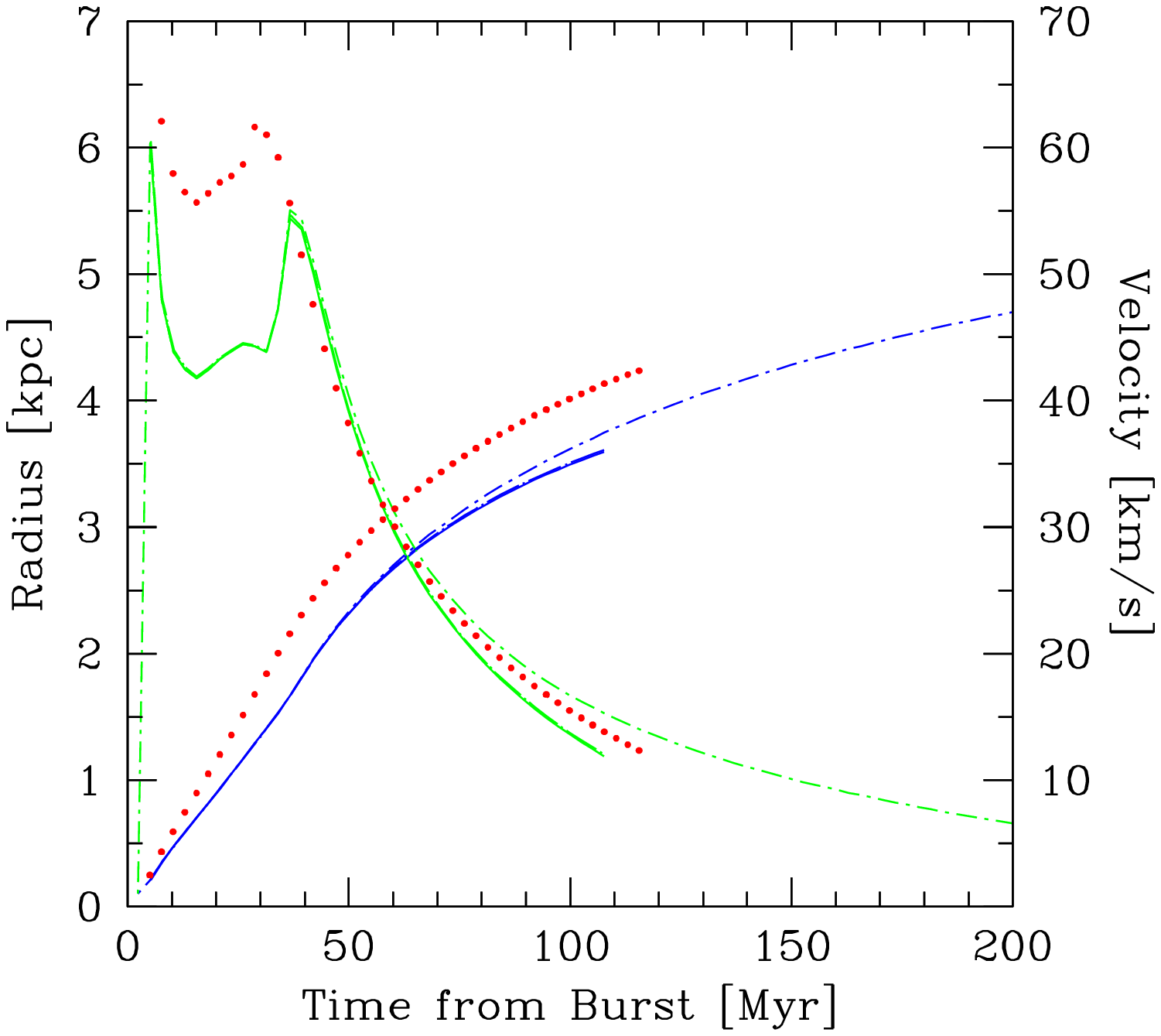}
\caption{Evolution of the shell radius and velocity in the fiducial case 
$f_\star=1$\%, $M_l=0.1\,\msun$ ({\it solid curves} in both panels, repeated for 
clarity). All other curves show solutions with the same parameters when in turn: 
cooling is inhibited ({\it small--dotted curves}), the mechanical luminosity is kept 
at the constant value $L_{38}=10$ ({\it short--dashed curves}), gravity is 
neglected ({\it long--dashed curves}), the IGM external pressure is set equal to 
zero ({\it dot--dashed curves}), and the halo gas mass is 50\% of the standard 
value ({\it large--dotted curves}). A case with thermal conductivity suppressed by 
two orders of magnitude cannot be distinguished from the solid curve.
\label{fig8}}
\end{figure}

The evolutionary equations (\ref{tsh}), (\ref{tsh1}), (\ref{Tevol})
have been integrated numerically. We start by analyzing the low efficiency case, 
$f_\star =1$\%, in some detail to highlight the relevant physics. As shown in 
Figure \ref{fig8}, the radius of the SN-driven bubble increases
with time up to a final stalling value of $R_f=3.5$~kpc, when pressure 
equilibrium is achieved. This happens in approximately 180 Myr, less than half
of the then Hubble time; up to this point the evolution is then largely unaffected 
by the Hubble expansion. In the inital stages we find $R_s\propto t^{1.1}$, 
a faster growth than given in (\ref{anal}), both because of the acceleration 
occurring in the halo density stratification and, less significantly, 
of the time--dependent mechanical luminosity. In the late stages of the evolution 
the shell evolves according to the `snowplough' (momentum conserving) solution,
$R_s \propto t^{1/4}$, as expected in the uniform IGM density field.
Gas will be lost from the halo if its specific enthalpy exceeds its gravitational
binding energy per unit mass. 
The distinctive feature of blow--away can be seen in the velocity profile
as a sudden jump of 10--15 km~s$^{-1}$ occurring at $t=40$ Myr. 
In all cases the velocity just before reacceleration is slightly lower than 
the escape velocity at the virial radius, $v_e(r_\vir)=48$~km~s$^{-1}$.
The outer shock is radiative as long as it runs into the ISM of the protogalaxy.  
After breakout ($t=40\,$Myr) the inverse Compton cooling time is 230 Myr 
and the shock becomes adiabatic. (Compton cooling dominates since we assume
the IGM to be photoionized, and collisional excitation cooling of H and He
is accordingly suppressed.)
For illustrative purposes, we also show solutions where the gravity term was 
turned off, the pressure of the IGM was set to zero, and 
the halo gas mass was set to 50\% of the standard value to mimic the possible 
presence of a cold galactic disk, which the (bipolar) outflow hardly couples to 
in numerical simulations (MacLow \& Ferrara 1999). It is only in the last case 
that 
the final stalling radius differs significantly from our fiducial solution.
Note that, except in the case of zero external pressure, we actually stop 
the numerical 
integration of the equation
at Mach number ${\cal M}=1$, as the shock decays into sound waves. We define the
radius where that happens as the `stalling radius', even if at that point the 
bubble pressure is still slightly higher than the external pressure of the IGM
and the expansion will continue for a while longer.
In practice, since the thin shell approximation assumes a strong shock, our 
solution actually breaks down before ${\cal M}=1$.

One feature of our model for the evolution of a wind--driven bubble is that
conductivity erodes the inside of the cold shell and thereby mixes some swept--up  
material into the hot cavity. If the SN ejecta carry magnetic fields, then
conductivity will be markedly reduced perpendicular to ${\bf B}$ and this process
will be inhibited. To check whether our solution depends crucially
on the value of the thermal conduction coefficient, we have run a case in which 
$\eta$ in equation (30) was decreased by a factor of 100.   
We find that the temperature profile of the gas within the cavity is increased
by less than $25\%$. Such a change leaves the rate of expansion of the cold 
shell essentially unaffected by the lower conductivity. This is because the 
(thermo)dynamics of the bubble is largely determined by the mechanical energy input 
and adiabatic expansion. 

\subsection{Varying the star formation efficiency}

\begin{figure}
\centerline{
\psfig{figure=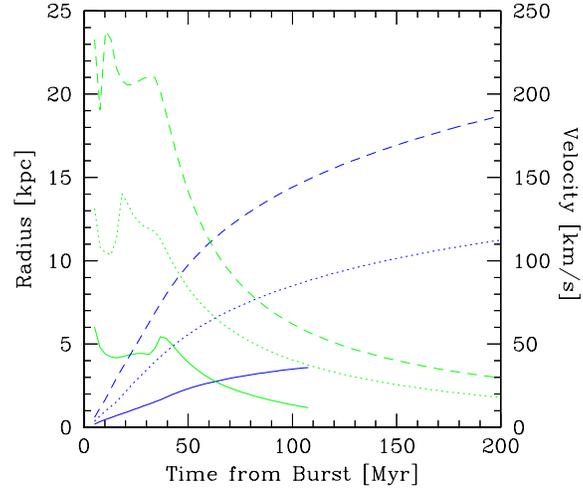,width=3.8in}}
\caption{Same as Fig. \ref{fig8}, but for the cases ($f_\star, M_l)=$ 
$(1\%, 0.1\,\msun)$  ({\it solid curves}, repeated for comparison), 
$(10\%, 0.1\,\msun)$ ({\it dotted curves}), and $(50\%, 0.1\,\msun)$  ({\it 
dashed curves}).
\label{fig9}}
\end{figure}

We now explore the effect of varying the star formation efficiency 
$f_\star$. The results of the numerical calculations are shown in Figure 
\ref{fig9}. The evolution is qualitatively similar in the three cases, but the 
final radius is significantly larger as $f_\star$ increases: for example, 
the case $f_\star = 10$\% produces a bubble radius which is 3.3 times larger 
than for $f_\star = 1$\%. Note that now the stalling point is reached only 
at ages comparable to the then Hubble time $t_H$, where our Newtonian solution 
breaks down. In the extreme case of $f_\star=50$\% the shell is still
expanding at a velocity of 30 km~s$^{-1}$ at $t=0.5\,t_H=200\,$Myr. 

\subsection{Varying the IMF}     

\begin{figure}
\centerline{
\psfig{figure=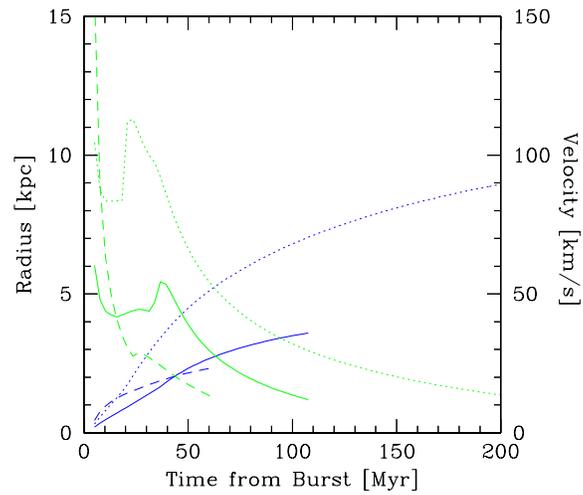,width=3.8in}}
\caption{Same as Fig. \ref{fig8}, but for the cases ($f_\star, M_l)=$ 
$(1\%, 0.1\,\msun)$  ({\it solid curves}, repeated for comparison), 
$(1\%, 5\,\msun)$ ({\it dotted curves}), and $(1\%, 30\,\msun)$  ({\it 
dashed curves}).
\label{fig10}}
\end{figure}

Variations in the lower mass cut--off of the adopted Salpeter IMF will
also affect our solution. Recall that the number of SNe produced by our 
three different choices of $M_l=(0.1, 5, 30)\,\msun$ is ${\cal N}=(1126, 6076, 
3000)$, respectively, for $f_\star=1\%$; obviously,
in the last case the progenitor stars are much more massive. The energy released
into the ISM by these very high mass SNe may well be lower than assumed here,
as part of the ejecta may fall back to form a black hole.
Even if this is not the case, the ability of very massive SNe to drive outflows
is limited because of their short lifetime, as shown in Figure \ref{fig10}. 
The moderately top--heavy IMF case, $M_l=5\,\msun$, generates a final radius of
about 10 kpc, \ie 2.5 times larger than for $M_l=0.1\,\msun$. For the very 
top--heavy IMF with $M_l=30\,\msun$, the shell barely reaches 2 kpc, 
in spite of being driven by 2.7 times more SNe than in the fiducial scenario. The 
reason is that in this case the pulsed mass and energy input only lasts for 
$\approx 3\,$Myr, and the increased peak mechanical luminosity ($L_{38}=10^3$) 
is not enough to counterbalance the strong external pressure of the dense halo 
gas: the bubble enters the `snowplough' phase well before blowing out into the 
IGM, and decelerates 
rapidly. To lift the halo gas out of the potential well it is then 
important that the energy 
injection continues at least until blow--away occurs; after that, the shell
will propagate to large distances while conserving momentum in the rarefied IGM.

\subsection{Bubble temperature}           

\begin{figure}
\centerline{
\psfig{figure=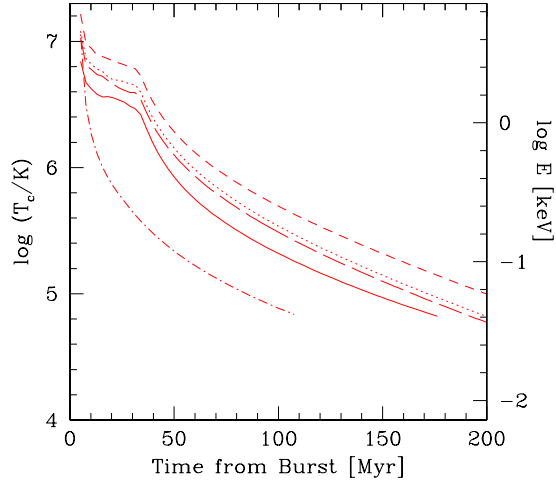,width=3.8in}}
\caption{Evolution of the gas temperature at the center of the expanding 
bubble for the five cases ($f_\star, M_l)=$ 
$(1\%, 0.1\,\msun)$  ({\it solid curve}), 
$(1\%, 5\,\msun)$  ({\it long--dashed curve}), 
$(1\%, 30\,\msun)$  ({\it dot--dashed curve}), 
$(10\%, 0.1\,\msun)$  ({\it dotted curve}), and 
$(50\%, 0.1\,\msun)$  ({\it short--dashed curve}). 
\label{fig11}}
\end{figure}

The evolution of the gas temperature $T_c$ at the center of the hot, metal--enriched
expanding bubble [recall that $T(r)=T_c (1-r/R_s)^{2/5}$ inside the cavity] 
is shown in Figure \ref{fig11}. While a fraction of order 
10\% of the then Hubble time is spent at high temperatures ($\log T_c= 6.6-7.2$), 
after blow--away cooling sets in very rapidly. This behavior is common to all 
cases except the most extreme one with $M_l=30\,\msun$, which essentially 
lacks the first very hot phase. Bubbles that do not stall continue to expand 
and cool. Inverse Compton cooling is the dominant source of energy loss as can 
be inferred from the slope of the temperature profile, which is much steeper 
than expected from pure adiabatic expansion (in this phase $R_s \propto t^{1/4}$, 
hence $T_c\propto t^{-1/2}$ if the gas is adiabatic). Thus, independently on 
whether a stalling radius is reached or not, all bubbles will continue to cool 
according essentially to the same law. 

\section{Discussion}

In the previous section we have modeled the evolution of SN--driven bubbles 
as they blow out from subgalactic halos and propagate into the intergalactic 
medium prior to the reionization epoch. We have seen that SN ejecta will 
escape the grasp of halos with virial temperatures $T_\vir\gta 10^{4.3}\,$K at 
$z=9$ (when they collapse from 2--$\sigma$ 
fluctuations) if a significant fraction of their 
baryonic mass can be converted into stars over a dynamical timescale. Depending
on the star formation efficiency, IMF, and detailed physics of the expansion,
we find that after about half of the then Hubble time these outflows will 
have produced a warm, 
$10^5$~K metal--enriched low--density region, with a size between a few and
several ten times larger than the virial radius of 
the source halo, surrounded by a colder,
dense shell which contains most of the swept--up mass.  
At $t=10^8\,$yr, typical shell bulk velocities range between 
$12\,\kms$ (for $f_\star=1\%$) to $40\,\kms$ ($f_\star=10\%$) and
up to $60\,\kms$ ($f_\star=50\%$). If the fraction of ionizing photons which escape 
the dense sites of star 
formation into the intergalactic space is greater than a few percent, SN--driven 
bubbles will propagate into a pre--photoionized medium, and their expansion 
will be halted by external pressure, gravity, and radiative losses. 
When the velocity of the outflow becomes subsonic, the enriched 
material gets dispersed by the random velocities in the IGM and the bubble looses
its identity. 

Knowing the mass within the bubble and the metal yield per 
supernova we can now derive the metallicity volution of the bubble.   
We have used the compilation of Todini \& Ferrara (2000) --essentially based 
on the Case A of Woosley \& Weaver (1995) -- who give the 
chemical composition of the ejecta for zero metallicity Type II SNe
(at redshift 9 the universe may be too young to host Type Ia SNe) 
as a function of their mass. At each time step
of the simulation, we check for the massive stars that went supernovae; we
sum over the mass of the heavy elements produced by them and divide by 
the mass of the hot gas within the bubble at that time. 
This is the typical metallicity inside the bubble if metals can be efficiently 
mixed within the cavity. Generally speaking, the metal--rich ejecta will 
be separated from the shell--evaporated 
gas by a contact discontinuity. The main  problem then is to disrupt this 
surface. When the cooling time in the cavity becomes shorter than its age,
a cooled shell will form which rapidly becomes unstable 
(Gull 1973) to Rayleigh--Taylor (RT) and Kelvin--Helmoltz (KH) instabilities.
The `fingers' produced by RT will penetrate deep into the high metallicity gas 
and may be eroded by KH instabilities due to the passage of rapidly moving hot 
gas, i.e. a mixing layer, perhaps creating  a rather uniform distribution of 
metals inside the bubble.

Further mixing, this time between the shell and the bubble gas, may occur 
after the expansion stalls. At that time, due to the restoring gravitational
force due $g$ on the shell from the DM halo, the acceleration vector points 
from the dense shell to the rarefied bubble, a RT unstable situation. 
The growth time of the RT instability on spatial scales $\lambda_s$ is
\begin{equation}
\label{RTgrow}
t_{\rm RT} \simeq \left[{{2\pi g \over \lambda_s } {(\Delta - 1)\over (\Delta +
1)}}\right]^{-1/2} \simeq \left({2\pi g \over \lambda_s }\right)^{-1/2}.
\end{equation}
If the density contrast between the shell and the hot cavity gas is $\Delta
\equiv n_s/n_h \gg 1$, and ${g} = 3 \times 10^{-11}$ cm s$^{-2}$
at $r=10$~kpc, $t_{\rm RT}$ is less than 10 Myr for any reasonable 
thickness (tens of parsecs) of the shell. Hence mixing of shell material
with the interior metal enriched gas may be quite rapid.
Ferrara, Pettini, \& Shchekinov (2000) have recently 
pointed out that additional processes related to mixing of the
gas caused by time--varying gravitational tidal forces produced by halo 
interactions must be invoked if 
the presence of metals at lower redshifts is found to be ubiquitous. 
The time scale of the process is difficult to estimate as it depends
on the peculariar velocities of the halos, which are predicted to be
low at high redshift. 
Note that the metal enrichment of the hot gas inside the bubble will
increase its cooling rate, which may become comparable with 
inverse Compton cooling. This is achieved around metallicity  
$\approx 0.5 Z_\odot$ at $z=9$. While our treatment may then underestimate the
cooling rate, we have seen in the previous section that cooling has a 
weak effect on the 
overall dynamics of the ejecta.

\begin{figure}
\centerline{
\psfig{figure=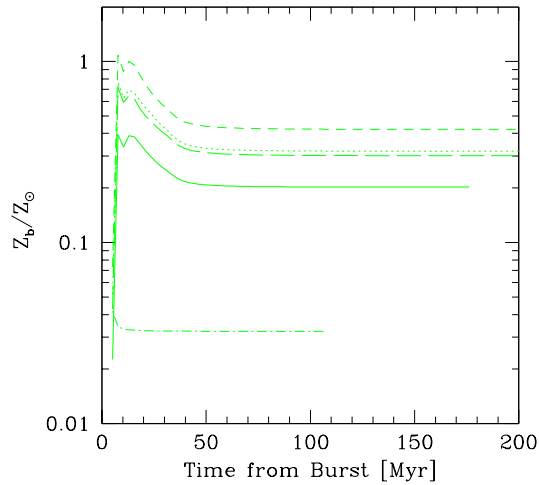,width=3.8in}}
\caption{Average metallicity of the bubble $Z_b$ (in units of solar) for 
the five cases shown in Fig. \ref{fig11} as a function of time from the burst. If 
heavy elements get mixed up with 
shell material, the mean metallicity of the shell--bubble outflow will
be two orders of magnitude lower than $Z_b$.
\label{fig12}}
\end{figure}

The time evolution of the bubble average metallicity, $Z_b$, is shown in 
Figure \ref{fig12} for the five cases studied. Its value is 
affected by the dynamical evolution of the bubble through the injection by 
evaporative of mass from the shell into the cavity. At later times, after
blow--away, $Z_b$ rapidly approaches a constant value determined by the
fact that (i) metal production ceases with the last explosion after 
$t=35$ Myr, and (ii) the evaporative mass flow rate (\ie the mass input in the
bubble) drops,
being extremely sensitive ($\propto T^{5/2}$) to the (rapidly decreasing)
temperature. The metallicity becomes then the ratio of two constants, and 
is determined by the past dynamical evolution of the outflow.
The final values are in the narrow range $Z_b = 0.2-0.4\,Z_\odot$, with
the only exception of the peculiar case with $M_l=30\,\msun$, which has 
$Z_b=3\times 10^{-2}\,Z_\odot$.
In Table 1 we show the expected bubble and shell masses at stalling. 
The final ratio $M_b/M_s$ is of order $1$\%.
Thus, if the heavy elements were completely mixed into the swept--up shell, 
then the resultant metallicities would
be lower than those derived above by about two orders of magnitude. It is possible, 
though, that the heavy 
elements may remain restricted to a region with a modest volume filling factor, 
and would not perhaps penetrate into the underdense 
medium between the halos. As already mentioned, the outcome depends on the onset 
of RT instabilities: 
non-linear development of these instabilities would lead to more widespread 
dispersal of the heavy elements through the IGM.  The details  are 
sensitive to the sound speed within the bubble when it stalls: this speed 
depends on how much small--scale internal mixing there has been within the 
bubble, due to conductivity, etc; but it could be substantially higher than 
the escape velocity. If so, then pressure--driven `fingers' of hot enriched 
material (originating within the bubble) could propagate out into the IGM, 
readily reaching distances of order the interhalo spacing if they 
maintained their identity and did not mix too rapidly. The effectiveness of 
mixing  would depend on the extent to which 
conductivity is inhibited by magnetic fields. It is worth noting that the 
bubble material (consisting substantially of supernova remnants) may well 
contain a magnetic field. If so, the dispersal of the first heavy elements 
and of the first (`seed') magnetic fields are intimately connected.  If 
the `fingers' are magnetised, then they may well propagate a long way 
before mixing with their surroundings.  Whether the intergalactic heavy elements 
are fully mixed is, however, an interesting  question that is still open. The 
\lya forest observations imply that most lines of sight through each `cloud' or 
filament encounter some heavy elements. But it is not clear that the mixing 
is uniform. It is possible that, even at redshifts of 3--5, the IGM is 
inhomogeneous on very small scales. Even in the regions that have been 
enriched, the metals could be  restricted  to magnetised `streaks' with 
a volume filling factor of as little as one percent. These would  corresponding to 
the `fingers' which we envisage having formed at $z=10$, which would have 
been sheared and distorted by the subsequent gravitational clustering that 
led to the fully--developed forest at $z=3-5$.

\begin{figure}
\centerline{
\psfig{figure=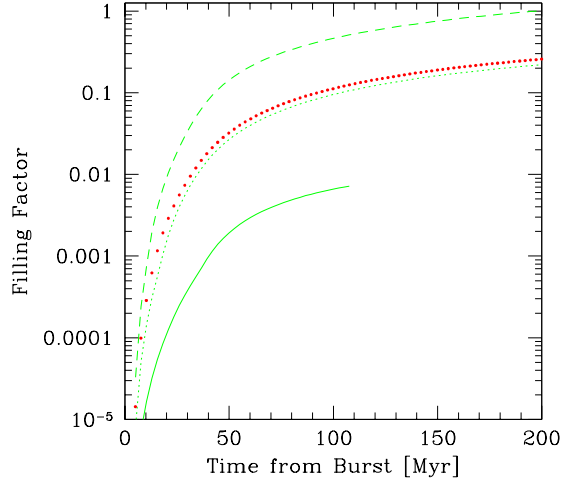,width=3.8in}}
\caption{Filling factor (porosity) of the metal enriched gas for the cases 
$(1\%, 0.1\,\msun)$  ({\it solid curve}), 
$(10\%, 0.1\,\msun)$  ({\it dotted curve}), 
$(10\%, 0.1\,\msun)$ where the halo gas mass is 50\% of the standard
value ({\it large--dotted curves}), and
$(50\%, 0.1\,\msun)$  ({\it short--dashed curve}). 
\label{fig15}}
\end{figure}

What is the spatial extent of the ensemble of wind--driven ejecta?  
In the adopted flat cosmology ($\Omega_M=1$, $h=0.5$, $\sigma_{8}=0.63$, 
$n=1$, $\Omega_bh^2=0.019$) and according to the Press--Schechter formalism, the 
comoving abundance of collapsed dark halos with masses $M\approx 10^8\,h^{-1}\,
\msun$ at $z=9$ is about $80\,h^3\,$Mpc$^{-3}$, corresponding to a mean 
proper distance between neighboring halos of $15\,h^{-1}\,$kpc
and to a total mass density parameter of order 3 percent. 
In Figure \ref{fig15} we plot the expected overall filling factor (or porosity) of
metal--enriched material for different scenarios, assuming a `synchronized' 
(over timescales shorter than the then Hubble time) population of starbursting halos.
As already mentioned,
with a star formation efficiency of $f_\star=10\%$ only a small fraction, about 
4\% percent, of the stars seen today would form at these early epochs. Still, 
the impact of such an era of pregalactic outflows could be quite significant, as
the product of stellar nucleosynthesis would be distributed over distance that are 
comparable to the mean proper distance between neighboring low--mass systems, 
i.e. filling factors $\gta 20\%$.
Larger filling factors may be obtained for larger efficiencies, moderately
top-heavy IMFs, halos where a significant fraction of the gas is 
in a galactic disk and does not couple to the outflow (as matter is ejected 
perpendicularly to the disk), 
or from a population of more numerous sources -- which would therefore have
to originate from lower amplitude peaks -- but are not required 
by current observations. 
Ferrara, Pettini, \& Shchekinov (2000) have recently considered the 
process of metal enrichment of the IGM, so it is useful to compare our main 
results with theirs. They determined the final radius of the bubbles by 
assuming that once blow--away has occurred, the shell enters the snowplough 
phase and it is finally confined by the IGM pressure. Essentially, their
stalling radius is much smaller than derived in this work because they
assumed a low cooling/star formation efficiency factor, $f_bf_\star=0.0025$
(corresponding for a halo of total mass $2\times 10^8\,\msun$ to a star 
formation rate of $10^{-3}\,\msun\,$yr$^{-1}$ over 30 Myr), \ie 
they studied outflows in a `quiescent' star formation mode 
that is perhaps more suitable to larger systems, rather than to the starbursting 
subgalactic halos we have focused on in this work.

\begin{figure}
\centerline{
\psfig{figure=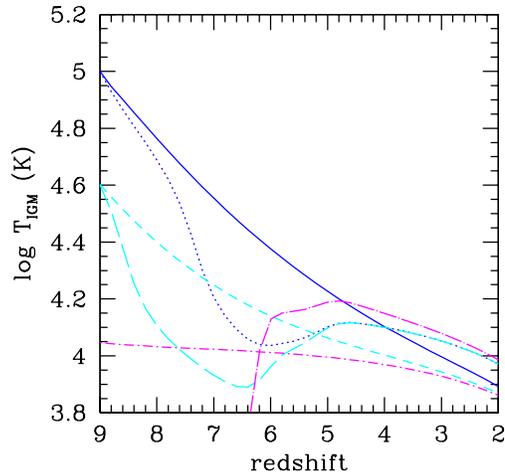,width=3.8in}}
\caption{Thermal history of intergalactic gas at the mean density in an 
Einstein--de Sitter universe with $\Omega_bh^2=0.019$ and $h=0.5$.
{\it Short dash--dotted line:} temperature evolution when the 
only heating source is a constant ultraviolet (CUV) background of intensity 
$10^{-22}\,\uvunits$ 
at 1 Ryd and power--law spectrum with energy slope $\alpha=1$. 
{\it Long dash--dotted line:} same for the time--dependent quasar ionizing 
background as computed by Haardt \& Madau (1996; HM).
{\it Short dashed line:} heating due to a CUV background  but with an initial 
temperature of $4\times 10^4\,$K at $z=9$ as expected from an early era 
of pregalactic outflows. 
{\it Long dashed line:} same but for a HM background. {\it Solid line:} heating 
due to a CUV background  but with an initial temperature of $10^5\,$K 
at $z=9$.
{\it Dotted line:} same but for a HM background.
\label{fig16}}
\end{figure}

It is interesting to note that, following an early era of SN--driven 
outflows, the enriched IGM may be heated on large scales 
to characteristic temperatures $T_\igm\gta 10^{4.6}\,$K if the star formation
efficiency is $\gta 10\%$, so as to `choke off' the 
collapse of further $M\lta 3\times 10^8\,h^{-1}\msun$ systems by raising 
the Jeans mass. In this sense the process may be self-regulating. 
While more detailed calculations -- which include the contribution to the IGM metal 
and heat content from all levels of the mass hierarchy as a function of 
cosmic time -- need to be done to fully assess the impact of subgalactic halos
on the thermal (and ionization) state of the IGM, we argue here that it is
unlikely that an early input of mechanical energy will be the dominant effect in
determining the ionization state of the IGM on large scales (cf. Tegmark 
\etal 1993). We have shown in \S\,3.5 that  
if the fraction of ionizing photons that escape the dense sites of star formation
into the intergalactic space is greater than a few percent, pregalactic outflows
will propagate into an IGM which has been pre--photoionized by the same massive 
stars which later explode as SNe. The
relative importance of photoionization versus shock ionization obviously depends
on the efficiency with which radiation and mechanical
energy actually escape into the intergalactic space. One can 
easily show, however, that, 
during the evolution of a  `typical' stellar population more
energy is actually lost in ultraviolet radiation than in
mechanical form (e.g. Madau 2000). This is because in nuclear burning 
from zero to solar metallicity ($Z_\odot=0.02$), the energy radiated per
baryon is $0.02\times 0.007\times m_\nH c^2$;  
about one third of it goes into
H--ionizing photons. While the same massive stars that dominate the UV light
also explode as SNe, the mass fraction radiated in photons above 1 Ryd
is $4\times 10^{-5}$, ten times higher than the fraction
released in mechanical energy. Of course, a pre--photoionized universe at
$10^4\,$K, heated up to a higher temperature by SN--driven winds, will recombine 
more slowly.  
The expected thermal history of expanding intergalactic 
gas at the mean density and with low metallicity is plotted in Figure \ref{fig16} 
as a function of redshift for a number of illustrative cases.
The code we use includes the relevant cooling and
heating processes and follows the non--equilibrium evolution of hydrogen and
helium ionic species in a cosmological context. The gas is allowed to interact 
with the CMB through Compton cooling and either with a time--dependent QSO ionizing 
background as computed by Haardt \& Madau (1996) or with a 
time--independent metagalactic flux of intensity $10^{-22}\,\uvunits$ at 1 Ryd 
(and power--law spectrum with energy slope $\alpha=1$). The temperature of the 
medium at $z=9$ -- where we start our integration -- has been either 
computed self--consistently from photoheating or  
fixed to be in the range $10^{4.6}-10^5\,$K expected from SN--driven 
bubbles with significant filling factors. The various curves show that the 
temperature of the IGM at $z=3-4$ will retain little memory of  
an early era of pregalactic outflows.

\acknowledgments
\ni Support for this work was provided by NASA through ATP grant NAG5--4236
and grant AR--06337.10-94A from the Space Telescope Science Institute (P.M.), 
by a B. Rossi Visiting Fellowship at the Observatory of Arcetri (P.M.),
by the Center for Computational Physics at Tsukuba University (A.F.), and by 
the Royal Society (M.J.R.). A.F., P.M., and M.J.R. also acknowledge 
the support of the EC RTN network ``The Physics of the Intergalactic
Medium''.

\begin{deluxetable}{lccrcc}
\tablewidth{0pt}
\tablecaption{Bubble and shell masses at stalling radius}
\tablehead{
\colhead{$f_\star$} & \colhead{$M_{\ell}/\msun$} & \colhead{$M_b/\msun$} & \colhead{$M_s/\msun$}  & \colhead{$M_b/M_s$}
}
\startdata
1\%       & 0.1        &$1.0 \times 10^5$&$1.6\times 10^7$& 0.7\%    \\
10\%       & 0.1        &$6.8 \times 10^5$&$5.5\times 10^7$& 1.2\%    \\
50\%       & 0.1        &$2.5 \times 10^6$&$1.6\times 10^8$& 1.5\%$^\dag$\\ 
1\%       & 5          &$4.0 \times 10^5$&$3.2\times 10^7$& 1.3\%    \\
1\%       & 30         &$1.0 \times 10^5$&$1.5\times 10^7$& 0.7\%    \\
\enddata
\tablecomments{$\dag$ Evaluated at $t=200\,$Myr.}
\end{deluxetable}


\end{document}